\documentstyle[12pt]{article}
\def\be{\begin{equation}}
\newcommand{\ee}{\end{equation}}
\def\ba{\begin{array}}
\def\ea{\end{array}}

\def\ni{\noindent}
\newcommand{\dsp}{\displaystyle}
\newcommand{\scs}[1]{{\scriptscriptstyle #1}}

\newcommand{\tr}{{\rm tr}\,}
\newcommand{\eq}[1]{eq.(\ref{#1})}
\newcommand{\eqr}[1]{(\ref{#1})}
\newcommand{\pvint}{\int\!\!\!\!\!\! - }
\renewcommand{\d}{\partial}

\newcommand{\R}{\scs R}
\newcommand{\N}{\scs N}
\newcommand{\M}{\scs M}
\renewcommand{\S}{\scs S}
\def\sp{\;\;\;\;\;\;}
\def\si{\mbox{self-intersection} }
\def\sis{\mbox{self-intersections} }
\def\ir{\mbox{irreducible representation} }

\def\A{{\cal A}}
\def\a{\alpha}
\def\b{\beta}

\def\hR{\hat{R}}
\begin{document}
\flushright{Phys. Lett. {\bf B329} (1994) 338}
\vskip2.5cm
\begin{center}
{\huge \bf Wilson loops in large N QCD on a sphere}\\
\vskip2cm
{\Large B. Rusakov} \\
\vskip.3cm
          {\em School of Physics and Astronomy \\
          Raymond and Beverly Sackler Faculty of Exact Sciences \\
          Tel-Aviv University, Tel-Aviv 69978, Israel }\\
\vskip.2cm
and
\vskip.2cm
          {\em ICTP, P.O.Box 586\\I-34014 Trieste, Italy }\\
\vskip2.5cm
{\large\bf Abstract.}
\end{center}
Wilson loop averages of pure gauge QCD at large N on a sphere are
calculated by means of Makeenko-Migdal loop equation.

\newpage

The exact solvability of QCD$_2$ was noticed long time ago by
A.A.Migdal \cite{Mig}. To realize it, one can integrate over $U(N)$ link
variables using only the orthogonality of group characters.

Following this, an exact solution of QCD$_2$ on an arbitrary
2-manifold was obtained in ref.~\cite{Rus1}.

In particular, the partition function for a closed surface of a genus $g$
and of area $A$ (coupling constant absorbed into area) has the form:
\be
Z_g(A)=\sum_{R} d_R^{2(1-g)}\exp\Big(-\frac{A}{2N}C_R\Big)
\label{Zg=}
\ee
where $d_R$ is the dimension of an $R$-th \ir,
\be
d_R = \prod_{1\leq i < j \leq N} \Big(1+\frac{n_i-n_j}{j-i}\Big)
\label{dim}
\ee
$C_R$ is the value of the second Casimir operator,
\be
C_R = \sum_{i=1}^{N} n_i(n_i +N+1-2i)
\label{Casimir}
\ee
where $n_i$ are highest weight components of
$R\equiv\{
n_1\geq n_2 \geq \ldots \geq n_\N\}$.

The Wilson loop average has the form \cite{Rus1}:
\be
W_g(C)=\sum_{\R_1,...,\R_\M}\Phi_{\R_1...\R_\M}
\prod_{k=1}^M d_{\R_k}^{2(1-g_k)}
\exp\Big(-\frac{A_k}{2N}C_{\R_k}\Big)
\label{Wg=}
\ee
where $M$ is the number of windows, $A_k$'s are their areas, $g_k$
is the ``genus per window" and the coefficient $\Phi_{\R_1...\R_\M}$
is the $U(N)$ group factor related to a form of the given contour
\footnote{
In particular, for a contour without \sis, this factor is the product,
taken over all pairs of neighboring windows $<ij>$, of Wigner
coefficients $\Phi_{\R_i\R_j}$ (see \eq{Wigner} below).}
(see ref.~\cite{Rus1} for details).

\bigskip

Recently, the partition
function \eqr{Zg=} has been reexpanded over $1/N$ and at large N
it has been interpreted as a sum over branched coverings \cite{Gross}
\footnote{The similar interpretation has been known for the Wilson loop
operators as well \cite{Kos}.}.

In order to develop a large N technique appropriate for calculations,
say, of loop averages, it is tempting to rewrite
the partition function \eqr{Zg=} at large N
as a path integral over continuous Young tables
\cite{Rus2}.
In the paper of Douglas and Kazakov \cite{Kazakov}, in the
framework of this approach,
a third order phase transition on a sphere was found. Recently,
Boulatov has calculated the Wilson loop average for the simplest contour
\cite{Dima}.
The goal of present paper is the calculation of an arbitrary Wilson
loop average at large N on a sphere.

First, we repeat briefly the steps of \cite{Rus2}-\cite{Dima}.
In the spherical case, $g=0$, the sum becomes divergent at small areas.
Hence, in the large $N$ limit, a non-trivial saddle-point should exist.
Let us introduce the continuous function \cite{Rus2},
\be
h(x)=\lim_{N\to\infty} \frac{1}{N}\Big(i - \frac{N}{2} - n_i\Big);
\hspace{1cm}
x=\frac{i}{N}-\frac{1}{2}
\ee
Then the saddle-point equation takes the form
\be
\frac{A}{2}h=\pvint \frac{dy\rho(y)}{h-y}
\label{sp}
\ee
Here, \be \rho(h)=\frac{dx}{dh} \label{rho} \ee
obeys the inequality
\be \rho(h)\leq 1 \label{rho<1} \ee
If \eq{rho<1} is ignored, then the solution of \eq{sp} is the
semi-circle distribution,
\be \rho(h)=\frac{1}{\pi}\sqrt{A-\frac{A^2h^2}{4}} \;\;,
\label{rho_weak} \ee
which is valid for the areas $A<\pi^2$. For the areas
$A>\pi^2$, the inequality \eqr{rho<1} is crucial and the solution of
\eq{sp} is \cite{Kazakov}:
\be
\rho(h)=\left\{\ba{ll}
-\frac{2}{\pi ah}\sqrt{(a^2-h^2)(h^2-b^2)}
\Pi_1\Big(-\frac{b^2}{h^2},\frac{b}{a}\Big) & \mbox{ for $-a<h<-b$}\\
1 & \mbox{ for $-b<h<b$}\\
\frac{2}{\pi ah}\sqrt{(a^2-h^2)(h^2-b^2)}
\Pi_1\Big(-\frac{b^2}{h^2},\frac{b}{a}\Big) & \mbox{ for $b<h<a$}
\ea\right. \label{rho_strong}
\ee
where $\Pi_1(x,k)$ is the complete elliptic integral of the third kind
with the modulus $k=\frac{b}{a}$ and
parameters are to be determined from the equations
\be
a(2E-k'^2K)=1 \hspace{1.5cm} aA=4K
\ee
At the critical value, $A_c=\pi^2$, the third order phase transition
takes place.

\bigskip

Since the saddle-point solution is known in both phases, one can
calculate the Wilson loop averages \eqr{Wg=} at large N on a sphere.

The simplest Wilson loop on a sphere corresponds to
$M=2$ and $g_1=g_2={1\over 2}$ (see \eqr{Wg=}),
\be W(A_1,A_2)=\frac{1}{NZ_0(A)} \sum_{\R,\S} d_\R d_{\scs S}
\Phi_{\R\S} \exp(-\frac{A_1}{2N}C_R-\frac{A_2}{2N}C_S)
\label{W} \ee
where \be
\Phi_{\R\S}=\int dU \chi_\R(U)\chi_\S(U^\dagger)\tr(U)\label{Wigner}\ee
is the multiplicity of an \ir $S$ in the tensor product of $R$ with the
fundamental representation $f$; $A=A_1+A_2$ is the total area.

The expression for this quantity valid for both phases
is \cite{Dima}:
\be
W(A_1,A_2)= \oint_C \frac{dh}{2\pi i} e^{A_1h-f(h)}
= \oint_{\bar{C}} \frac{dh}{2\pi i} e^{-A_2h+f(h)}
\label{W=}
\ee
where
\be f(h) = \int \frac{dy\rho(y)}{h-y} \label{f} \ee
and the contour $C$ encircles the cut
of $f(h)$ (contour $\bar{C}$ encircles the cut in the opposite
direction). Obviously, $W(A_1,0)=W(0,A_2)=1$.

In the {\it weak coupling phase},
$f(h)=\frac{A}{2}h-\sqrt{\frac{A^2h^2}{4}-A}$,
and in terms of the new variable $z=\frac{\sqrt{A}}{if}$
the integral \eqr{W=} takes the form
\be
W_{\rm wc}(A_1,A_2)=\frac{1}{i\sqrt{A}}\oint_C \frac{dz}{2\pi i}
\Big(1+\frac{1}{z^2}\Big) e^{i\a z+i\frac{\b}{z}}
\ee
where $\a=\frac{A_1}{\sqrt{A}}$ and $\b=\frac{A_2}{\sqrt{A}}$.
Expanding the exponential and taking the residue at zero, one find
\be\ba{rl}
W_{\rm wc}(A_1,A_2)&=\frac{1}{i\sqrt{A}}{\dsp \oint_C} \frac{dz}{2\pi i}
\Big(1+\frac{1}{z^2}\Big) {\dsp \sum_{n=0}^{\infty} }
\frac{i^{2n+1}}{n!(n+1)!}(\a^{n+1}\b^nz+\a^n\b^{n+1}\frac{1}{z})
\vspace{1pc}
\\
&={\dsp \sum_{n=0}^{\infty} }
\frac{(-1)^n}{n!(n+1)!}(\a\b)^n=\sqrt{\frac{A_1+A_2}{A_1A_2}}
J_1\Big(\sqrt{\frac{4A_1A_2}{A_1+A_2}}\Big)\;\;\;. \ea \ee

In what follows we will also need $A\to\infty$ limit of \eqr{W=}. As
it has been shown in ref.~\cite{Dima}, the function \eqr{f} in this
case takes the form $f(x)=\log\frac{2x+1}{2x-1}$. For example, in the
case $A_1=\infty$, we have\be
W_{\rm sc}(\infty,A_2)=-\frac12\oint_O \frac{dx}{2\pi i}
\frac{x-1}{x+1}e^{\frac{A_2}{2}x}=e^{-\frac{A_2}{2}}
\ee
{\it i.e.}, standard QCD$_2$ area law.

In principle, all set of \eqr{Wg=} can be calculated in the same fashion.
However, since an expression for simple loop \eqr{W=} is known, all
others Wilson loops can be derived easily from the loop equation.

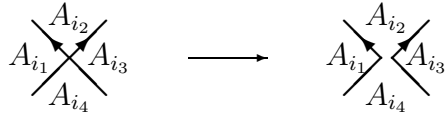
\begin{figure}
\centering
\begin{picture}(200,100)(- 100,- 50)
\thicklines
\put(-60, 0){\line(-1,1){14}}
\put(-60, 0){\vector(-1,1){8}}
\put(-60, 0){\line( 1,1){14}}
\put(-60, 0){\vector(1,1){8}}
\put(-60, 0){\line(1,-1){14}}
\put(-60, 0){\line(-1,-1){14}}

\put(58, 0){\line(-1,1){14}}
\put(58, 0){\vector(-1,1){8}}
\put(62, 0){\line( 1,1){14}}
\put(62, 0){\vector(1,1){8}}
\put(62, 0){\line(1,-1){14}}
\put(58, 0){\line(-1,-1){14}}

\put(-75, 0){\makebox(0,0){$A_{i_1}$}}
\put(-60,15){\makebox(0,0){$A_{i_2}$}}
\put(-45, 0){\makebox(0,0){$A_{i_3}$}}
\put(-60,-15){\makebox(0,0){$A_{i_4}$}}
\put( 45, 0){\makebox(0,0){$A_{i_1}$}}
\put( 60,15){\makebox(0,0){$A_{i_2}$}}
\put( 75, 0){\makebox(0,0){$A_{i_3}$}}
\put( 60,-15){\makebox(0,0){$A_{i_4}$}}

\thinlines
\put(- 15,0){\vector(1,0){30}}

\end{picture}

\caption[x]{\footnotesize Disconnection of a contour at the \si point}
\label{fig1}
\end{figure}

The equation for Wilson loops derived by Yu.Makeenko and A.Migdal a long
time ago \cite{MM}, has especially simple form in two dimensions
\cite{KK}:
\be \hR_i W_n(C)=W_q(C_i^1)W_{n-q-1}(C_i^2) \sp i=1,...,n \label{MMeq}\ee
where $n$ is a number of \sis of contour $C$.
Each equation of system \eqr{MMeq} corresponds to given $i$-th point of
\si of $C$ ($C_i^1$ and $C_i^2$ are results of
disconnection of contour $C$ in this point)\footnote{
Actually, on a sphere (as well as on any compact surface), the number of
independent equations of the system \eqr{MMeq} can be less, than $n$.
Nevertheless, it turns out that in the spherical case the
system \eqr{MMeq} is sufficient to derive a complete set
of loop averages.}.
Here, differential operator $\hR_i$ acting in $i$-th \si
point is the linear combination of derivatives with respect to areas of
windows $i_1,...,i_4$ neighboring at this point (in what follows, we
use the notation $\d_k\equiv \frac{\d}{\d A_k}$),
\be \hR_i = \d_{i_1}-\d_{i_2}+\d_{i_3}-\d_{i_4} \label{R=}\ee
where different signs correspond to windows having common boundary
\footnote{We do not fix the total sign here and below since it is always
can be restored from the condition $W(0,...,0)=1$.} (Fig.\ref{fig1}).

If all Wilson averages for loops with number of \si less, than $n$, are
known, then we know an explicit form of RHS of \eqr{MMeq}. Solving
recurrently the system \eqr{MMeq} we get all set of loop averages.

Starting point of \eqr{MMeq} is equation for {\bf 8-like contour}
drawn on Fig.2. In this case, the RHS of \eqr{MMeq} has
the form of product of two simple loops, $W_0$ (\eq{W=}), which
could not be derived from the Makeenko-Migdal equation.

The equation, corresponding to Fig.2, has the form
\be \hR W_1(A_1,A_2,A_3)=
\oint \frac{dx}{2\pi i}\oint\frac{dy}{2\pi i}
e^{A_1x+A_2y-f(x)-f(y)}\ee
where \be \hR = \d_1+\d_2-2\d_3 \ee
The natural anzatz is:
\be W_1(A_1,A_2,A_3)=
\oint \frac{dx}{2\pi i}\oint\frac{dy}{2\pi i}
e^{A_1x+A_2y-f(x)-f(y)}F\ee
Then, function $F=F(x,y;A_1,A_2,A_3)$ should satisfy the equation:
\be (\hR +x+y)F=1\;\;.\label{Feq}\ee
The solution of \eqr{Feq} is:
\be F=\frac{1+e^{-(x+y)(cA+\frac{A_3}{2})}
\xi(A_1,A_2,A_3)} {x+y} \ee
where function $\xi(A_1,A_2,A_3)$ should satisfy the equation:
\be \hR\xi=0 \ee
and $c$ is arbitrary constant.
An explicit form of function $\xi$ and value of constant $c$
should be defined from obvious conditions:
\be W_1(0,A_2,A_3)=W_0(A_2,A_3)\;\;,\ee
\be W_1(A_1,0,A_3)=W_0(A_1,A_3)\;\;,\ee
which fix unambiguously the function $F$. The result is:
\be W_1(A_1,A_2,A_3)=\oint \frac{dx}{2\pi i}\oint\frac{dy}{2\pi i}
\frac{e^{A_1x+A_2y-f(x)-f(y)}} {x+y} \label{Wo}\ee
Let us notice that this expression can be equally rewritten as
\be W_1(A_1,A_2,A_3)=P\oint \frac{dx}{2\pi i}\oint\frac{dy}{2\pi i}
\frac{e^{A_1x+(A_1+A_3)y-f(x)-f(y)}} {x-y} \label{WWo}\ee
where $P$ means that integration over $x$ should be performed after
integration over $y$. The equivalence between \eqr{Wo} and \eqr{WWo}
corresponds to the fact that the contour drawn on Fig.2
coincides with one drawn on Fig.3.

The known answers for a plane \cite{KK}-\cite{Kaz} are follow from
\eqr{Wo} and \eqr{WWo} as well:
\be W_1(A_1,A_2,\infty)= e^{-\frac{A_1+A_2}{2}}\label{A_3=inf}\ee
\be W_1(A_1,\infty,A_3)= e^{-A_1-\frac{A_3}{2}}(1-A_1)\label{A_2=inf}\ee
\be W_1(\infty,A_2,A_3)= e^{-A_2-\frac{A_3}{2}}(1-A_2)\label{A_1=inf}
\;\;\;.\ee

\bigskip

In the case of {\bf two self-intersections} there are two different
types of contours (Fig.4).

Since $W_0$ and $W_1$ are known,
we are able to write down explicitly an equation for $W_2$'s.
For example, for contour Fig.4~(a), we have
\be
\hR_1 W_2(A_1,A_2,A_3,A_4)=W_0(A_1,A_2+A_3+A_4)W_1(A_2,A_3,A_1+A_4) \ee
\be
\hR_2 W_2(A_1,A_2,A_3,A_4)=W_0(A_3,A_1+A_2+A_4)W_1(A_1,A_2,A_3+A_4) \ee
where \be \hR_1= \d_1+\d_2-2\d_4 \;\;,\ee
      \be \hR_2= \d_2+\d_3-2\d_4 \;\;.\ee

Repeating the steps of previous case, we derive
\be W_2(A_1,...,A_4)=\oint\frac{dx_1}{2\pi i}e^{-f(x_1)}
...\oint\frac{dx_3}{2\pi i}e^{-f(x_3)}
\frac{e^{A_1x_1+A_2x_2+A_3x_3}}{(x_1+x_2)(x_2+x_3)}\label{two_a}\ee
while for contour Fig.4~(b), the answer is
\be W_2(A_1,...,A_4)=\oint\frac{dx_1}{2\pi i}e^{-f(x_1)}
...\oint\frac{dx_3}{2\pi i}e^{-f(x_3)}
\frac{e^{A_1x_1+(A_2+A_3)x_2+A_3x_3}}{(x_1+x_2)(x_3-x_2)}\label{two_b}\ee

It is also instructive to consider one more example,
the contours where pair (or pairs) of windows are separated by more than
one \si point. The first example of such a contour we match while
consider the case of {\bf three self-intersections}. The contour drawn
on the Fig.5 has two windows, 1 and 4, separated by three
\si points.
In this case, the solution of loop equation is
\be\ba{rl}
&W_3(A_1,...,A_5) =\vspace{1pc}
\\
&\oint\frac{dx}{2\pi i}\oint\frac{dy}{2\pi i}e^{-f(x)-f(y)}
\frac{e^{A_1x+A_4y}}{x+y}(e^{A_2(x+y)}+e^{A_3(x+y)}-e^{(A_2+A_3)(x+y)})
\ea\label{W3=}\ee
Let us notice that the structure of expression
under integral in \eqr{W3=} is in one to one correspondence to structure
of the plane solution \cite{KK} which we easily reproduce from \eqr{W3=}:
\be W_3(A_1,...,A_4,\infty)=
e^{-\frac{A_1+A_4}{2}}(e^{-A_2}+e^{-A_3}-e^{-A_2-A_3})
\label{W3_plane}\ee

\bigskip

Now, following previous consideration, we are able to formulate {\bf
general solution} for arbitrary $W_n(C)$:
\be W_n(C)=P\oint\frac{dx_1}{2\pi i}...\oint\frac{dx_m}{2\pi i}
\prod_{k=1}^m e^{\A_kx_k-f(x_k)}\prod_{<ij>}\frac{1}{x_i\pm x_j}
\sum_{q=1}^{n_{ij}>1}(-1)^q e^{\A_q^{ij}(x_i\pm x_j)}
\label{GEN}\ee
where: $m$ is number of windows of the {\em disk} topology appearing
as a result of disconnection of $C$ at all \si points
(we associate variables $x_k$ with all such a windows)
\footnote{Though choice of $x_k$ is not unambiguous (we can attach it to
``internal" or to ``external" regions of the window), it doesn't lead to
ambiguity in the answer.}; $\A_k$ is an area of such a window\footnote{
It can be equal to sum of areas $A_k$ of
several original windows as we have shown in \eqr{WWo} and \eqr{two_b}.};
the second product goes over all pairs $<ij>$ of windows from this set
which are separated by (at least one) \si point; minus sign of
``propagator" $\frac{1}{x_i\pm x_j}$ corresponds to the case of
self-overlapping, when $i$-th window is internal with respect to $j$-th
one; $P$-symbol means that, in the latter case,
integration over $x_j$ should be performed before integration over $x_i$;
if $i$-th and $j$-th windows are separated by $n_{ij}>1$ \si points
\footnote{Actually, $n_{ij}$ is the odd number.}, then the sum goes
over all disconnections at such points and $\A_q^{ij}$ is the area of
the overlapping region appeared as result of $q$-th disconnection (see
\eq{W3=}).

\bigskip

Eq. \eqr{GEN} is the answer for arbitrary Wilson loop average in
large N pure QCD on a sphere. Let us emphasize that it
is valid in both weak and strong coupling phases. A complete information
about phase transition is contained in the function $\rho(x)$ ($f(x)$ as
well).
This function is given by \eq{rho_weak} in the weak coupling phase and by
\eq{rho_strong}
in the strong coupling phase.

\bigskip

\ni
{\large\bf Acknowledgments.}\\
\ni
I am grateful to D.Boulatov for fruitful discussions.
This research has been supported in part by the Basic Research Foundation
administered by the Israel Academy of Sciences and Humanities, by a
grant from the US$-$Israel Binational Science Foundation (BSF),
Jerusalem, Israel, and also by the Israel Ministry of Absorption.

\end{document}